\documentclass[12pt,aps,prd,floatfix,nofootinbib,a4paper,superscriptaddress]{revtex4-2}
\pdfoutput=1

\usepackage{amsmath, amssymb, amsfonts, amsthm, latexsym, epsfig, mathrsfs, xcolor, bbm, slashed, braket, thmtools}

\usepackage[all]{xy}
\usepackage{comment}

\usepackage[inline]{enumitem}

\usepackage{setspace}
\usepackage[marginal, multiple]{footmisc}

\usepackage[T1]{fontenc}
\usepackage[utf8]{inputenc}
\usepackage{lmodern}

\usepackage[colorlinks, allcolors=blue!70!black, linktocpage]{hyperref}

\numberwithin{equation}{section}

\usepackage{cleveref}
\usepackage{color}

\usepackage{microtype}

\usepackage{floatrow}
\let\OLDtableofcontents\tableofcontents
\renewcommand\tableofcontents[1]{%
    {\baselineskip 0.5ex %
	\OLDtableofcontents{#1}}%
}

\let\OLDthebibliography\thebibliography
\renewcommand\thebibliography[1]{%
	\setstretch{1.079} 
	\OLDthebibliography{#1}%
	\small %
	\setlength{\itemsep}{0.2\baselineskip} 
}

\let\OLDfootnote\footnote
\renewcommand\footnote[1]{%
	\setlength{\footnotesep}{0.75\baselineskip}%
	{\footnotesize \OLDfootnote{#1}}%
}

\setlist[enumerate]{noitemsep, label=(\arabic*), ref=(\arabic*)}

\renewcommand\thesection{\arabic{section}}
\renewcommand\thesubsection{\arabic{subsection}}

\makeatletter
\def\p@subsection{\thesection.}
\def\p@subsubsection{\thesection.\thesubsection.}
\makeatother 

\theoremstyle{plain}

\theoremstyle{definition}
\newtheorem{definition}{Definition}[section]

\creflabelformat{equation}{#2#1#3}

\crefname{section}{sec.}{sec.}
\crefname{appendix}{Appendix}{Appendices}
\crefname{figure}{Fig.}{Figs.}
\crefname{table}{Table}{Tables}

\crefname{definition}{Def.}{Defs.}
\crefname{prop}{Prop.}{Props.}
\crefname{lemma}{Lemma}{Lemmas}
\crefname{corollary}{Cor.}{Cors.}
\crefname{thm}{Theorem}{Theorems}
\crefname{remark}{Remark}{Remarks}

\crefname{ass}{Assumptions}{Assumptions}
\crefname{property}{Properties}{Properties}

\newcommand{\be}{\begin{equation}\begin{aligned}}
\newcommand{\ee}{\end{aligned}\end{equation}}

\newcommand{\mc}{\mathcal}

\newcommand{\ms}{\mathscr}

\newcommand{\bb}{\mathbb}

\newcommand{\eqsp}{\, ,\quad} 

\newcommand{\note}[1]{{\bf \textcolor{red!70!black}{#1}}} 

\newcommand{\Lie}{\pounds} 
\newcommand{\defn}{\mathrel{\mathop:}=} 

\newcommand{\hateq}{\mathrel{\mathop {\widehat=} }} 

\newcommand{\de}{\partial}

\let\oldint\int
\renewcommand{\int}{\oldint\limits}

\let\oldlim\lim
\renewcommand{\lim}{\oldlim\limits}

\renewcommand{\bar}{\overline}

\newcommand{\scri}{\ms I}

\newcommand{\pb}[1]{\underleftarrow{#1}}

\newcommand{\Hilb}{\mathscr{H}}

\newcommand{\antiHilb}{%
\hspace{4pt} 
  \vbox{%
    \hrule height 0.5pt
    \kern0.25ex
    \hbox{%
      \kern-0.3em
      \ifmmode\Hilb\else\ensuremath{\Hilb}\fi
      \kern0em
    }
  }
}

\begin{document}

\setstretch{1.2}

\title{On symmetries of gravitational on-shell boundary action at null infinity}

\author{Shivam Upadhyay}
\email{shivam.dep@gmail.com}
\affiliation{Chennai Mathematical Institute, Siruseri, SIPCOT, Chennai 603103}
\affiliation{Raman Research Institute, Sadashivnagar, Bengaluru, 560080, India}

\begin{abstract}
\begin{center} \textbf{Abstract}
\end{center}
We revisit the gravitational boundary action at null infinity of asymptotically flat spacetimes. We fix the corner ambiguities in the boundary action by using the constraint that (exponential of) the on-shell action leads to tree-level 5-point amplitude in eikonal approximation in a generic supertranslated vacuum. The subleading soft graviton theorem follows naturally from the on-shell action when the BMS group is extended to include superrotations. An infinite tower of Goldstone modes is proposed by `generalizing' the Geroch tensor to incorporate a set of divergence-free symmetric traceless tensors on the sphere. This generalization leads to $\text{sub}^{n}$-leading soft insertions in the tree-level $\mc{S}$ matrix, thus paving the way to understanding the infinite tower of tree-level symmetries within this framework.
\end{abstract}

\maketitle
\newpage 
\tableofcontents

\section{Introduction}\label{sec:intro}
Over the past decade, there have been several seminal developments in our understanding of the so-called infra-red triangle\cite{Strominger:2017zoo,Strominger:2014pwa}. Infra-red triangle is a conceptual edifice that connects three disparate ideas, namely low-frequency gravitational observables\cite{Braginsky:1987kwo,Christodoulou:1991cr,Bieri:2013hqa}, asymptotic symmetries\cite{Bondi:1962px,Sachs:1962zzb,Ashtekar:1981bq,Ashtekar:1987tt,Strominger:2013jfa}, and quantum soft theorems\cite{Weinberg:1965nx,He:2014laa}.  Through these interconnections, we now understand that the symmetry group of classical gravitational scattering \emph{contains} the well-known Bondi, Metzner, and Sachs (BMS) group which is a semi-direct product of the Lorentz group and supertranslations\cite{Bondi:1962px,Sachs:1962zzb}, with the complete asymptotic symmetry group still not known in classical general relativity\cite{Freidel:2021ytz,Choi:2024ygx}. The BMS group has been further extended to include superrotations\cite{Barnich:2009se,Campiglia:2014yka,Campiglia:2015yka,Kapec:2014opa}. The term superrotations collectively denotes the enlargement of Lorentz transformations to infinite-dimensional extensions generated either by meromorphic vector fields or smooth diffeomorphisms on the sphere. These two extensions along with supertranslations are known as extended BMS (eBMS) and generalized BMS (gBMS) respectively.

The situation is similar in QED. In this case, the analog of the BMS group is the Abelian group of large gauge transformations whose conservation law in any classical scattering generates the formula for what is known as, velocity memory.  It was shown in \cite{He:2014cra} that the Ward identity associated with this conservation law is simply the Weinberg soft photon theorem statement, thus generating an infra-red triangle in QED.  

The relationship between gravitational $\mc S$ matrix and BMS symmetries as conjectured by Strominger\cite{Strominger:2013jfa} mandates that any gravitational scattering amplitude must satisfy the supertranslation conservation laws. For generic loop amplitudes, this fact manifests itself into a statement about vacuum transitions between the past and the future. The difference between the two vacuua being quantified by the Weinberg soft factor (gravitational memory)\cite{He:2014cra,Choi:2017ylo,Prabhu:2022zcr}\footnote{Existence of such dressed states in gravity however is far more nuanced than in QED and we refer the reader to \cite{Prabhu:2022zcr,Prabhu:2024lmg,Prabhu:2024zwl} for details.}.  Tree-level amplitudes however can be computed using Feynman diagrammatics which do not directly refer to constraints imposed by asymptotic Ward identities. This is however not an issue as the supertranslation Ward identity is equivalent to Weinberg soft graviton theorem and this theorem is satisfied by a number of generic tree-level amplitudes in gravity. On the other hand, a certain class of tree-level amplitudes (e.g. amplitudes in the eikonal regime) can also be computed ``in position space'' using the on-shell boundary action\cite{Fabbrichesi:1993kz}. This perspective is of course not new and was pioneered by  
Arefeva, Faddeev, and Slavnov (AFS)\cite{Arefeva:1974jv} in 70s. This path integral formulation of the $\mc S$ matrix is interesting from the perspective of asymptotic symmetries as it relies on the specification of boundary data at future and past null infinities\footnote{In this paper we only consider massless matter states as external data for gravitational scattering.}. Fabbrichesi et al\cite{Fabbrichesi:1993kz} reproduced the 4-point tree level amplitude in the eikonal approximation from the on-shell boundary action when there is no gravitational memory. 

It is then a natural question to ask how the on-shell boundary action transforms under supertranslation symmetry and in what sense the modified action leads to an inelastic (tree-level) eikonal amplitude in the soft limit. Our work is closely related in spirit to the work of Kim, Kraus, Monten, and Myers\cite{Kim:2023qbl} in which they derived the soft photon theorem using the AFS $\mc{S}$ matrix for massless QED case. However there are subtle technical differences with regards to the definition of boundary action in asymptotically flat spacetimes with non-trivial memory mode turned on.

In this paper, we use the construction of Lehner, Myers, Poisson and Sorkin (LMPS) \cite{Lehner_2016} to construct an on-shell boundary action at null infinity in a generic supertranslated vacuum. We show that this action has the so-called corner term ambiguities\footnote{The corner term ambiguities have previously been noted in \cite{Jubb:2016qzt,Odak:2022ndm,Freidel:2021fxf}.}. The constraint that the (exponential of this action) leads to 5-point eikonal scattering in a supertranslated background fixes the ambiguities in the corner term. We compute the boundary terms in the conformal spacetime picture of Penrose et al.\cite{Penrose:1964ge,Geroch:1977big,Ashtekar:1981bq,Ashtekar:1987tt,Ashtekar:2014zsa,Ashtekar:2018lor} where null infinity exists as a codimension-1 null hypersurface. The corner ambiguities can be understood as arising from the residual conformal freedom, which does not change the metric \emph{on} the boundary. The radiative flux term plays an important role, and its covariantization leads to an extra corner contribution to the boundary action when working in an arbitrary conformal frame. This term corresponds to the subleading soft news smeared with the trace-free Geroch tensor. It is crucial to relate the AFS $\mc{S}$ matrix to the subleading soft graviton theorem when allowing the extension of BMS to include (holomorphic) superrotations. However, the flux term is only covariant up to a total divergence on the cross-section as shown in \cite{Rignon-Bret:2024gcx}. `Generalizing' the Geroch tensor by incorporating a set of divergence-free symmetric traceless tensors on the sphere (away from the singularities) leads to an infinite number of terms in the on-shell action, giving us a soft insertion at $O(\omega^{n-1})$.

The structure of this paper is as follows: In \Cref{sec:onshell}, we give a short review of the path integral approach to the $\mc{S}$ matrix and its application to eikonal gravitational scattering first done by Fabbrichesi et al\cite{Fabbrichesi:1993kz}. We next review the work of LMPS to derive an action with corner terms on null boundaries. In \Cref{sec:conformal frame}, using LMPS formulation we derive the boundary action at null infinity and give its expression in the Bondi coordinates. We then fix the corner ambiguities in the boundary action by using the constraint that the (exponential of) on-shell action leads to a 5-point amplitude with an outgoing soft graviton. In \Cref{sec:celestial}, we work in an arbitrary time-independent conformal frame and covariantize the flux term to get an extra contribution to the boundary action. This term is precisely what allows us to relate the AFS $\mc{S}$ matrix to the subleading soft graviton theorem when the Lorentz group is enhanced to include superrotations. In \Cref{sec:subsubleading}, we propose an infinite tower of Goldstone modes by generalizing the Geroch tensor to include a set of divergence-free symmetric traceless tensors on the sphere (away from the singularities) to show that the on-shell action hints towards an exposition of an infinite tower of tree-level soft theorems in this framework. We end with some concluding remarks and open questions.

\textbf{Conventions:} Quantities on the physical spacetime are denoted with a``hat'' $\hat{}$.  The ``$\hateq$'' denotes equality on $\ms{I}$. We use the lowercase Latin indices a,b,c,... for tensor fields in the spacetime and uppercase Latin indices A,B,C,... for the coordinates on the celestial sphere.
\section{On-shell boundary term in general relativity}\label{sec:onshell}
In this paper, we consider vacuum general relativity with vanishing cosmological constant in the setting of conformally completed asymptotically flat spacetimes. We restrict attention to configurations involving only massless excitations, so that the relevant asymptotic boundaries are future and past null infinity, $\mathscr{I}^\pm$. The boundary $\partial \mathcal M$ is therefore taken to consist of $\mathscr{I}^+ \cup \mathscr{I}^-$ (together with possible corner contributions), and the boundary data correspond to the radiative degrees of freedom of the gravitational field.

The path integral formulation of the $\mathcal{S}$-matrix can be expressed schematically as
\be
\mathcal{S}(p_{\text{in}}, p_{\text{out}}) = \int [\mathcal{D}\phi \,\mathcal{D}\hat g]\, e^{i S[\phi,\hat g]} ,
\ee
where the functional integral is taken over field configurations interpolating between prescribed asymptotic in and out states at null infinity. In the semiclassical (tree-level) limit, the path integral is dominated by classical solutions satisfying the appropriate boundary conditions, and the scattering amplitude can be extracted from the action evaluated on such solutions.  A key feature of general relativity is that the Einstein--Hilbert action does not admit a well-posed variational principle without the addition of boundary terms. The variation of the action takes the form
\be
\delta S = \int_{\mathcal M} E^{ab}\,\delta \hat g_{ab} + \int_{\partial \mathcal M} \theta(\hat g;\delta \hat g),
\ee
where $E^{ab}=0$ are the equations of motion and $\theta$ is the symplectic potential\footnote{In the language of forms, \(\theta\) is a \(d-1\) form.}. When the equations of motion are satisfied, the variation reduces entirely to a boundary contribution,
\be
\delta S \big|_{\text{on-shell}} = \int_{\partial \mathcal M} \theta .
\ee
For null boundaries, the pullback of the symplectic potential admits a decomposition of the form \cite{Chandrasekaran:2020wwn}
\be
\pb \theta = -\delta S_\mc N + d\beta + \Theta ,
\ee
where $S_\mc N$ is an integrable functional of the boundary data, $\beta$ encodes corner ambiguities, and $\Theta$ represents non-integrable flux contributions \footnote{\(d\) is the exterior derivative on the spacetime.}. It follows that
\be
\delta S = -\delta S_\mc N + \int_{\partial \mathcal M} (d\beta + \Theta) .
\ee
where the corner terms \(\beta \) localize to future and past boundaries of \(\partial \mc M \). Note that corner terms do not affect the variational principle since \(d^2=0 \). Thus, up to corner terms and flux contributions, the on-shell action is determined by the integrable functional $S_\mc N$ defined on the boundary. In this sense, the on-shell action plays the role of a Hamilton--Jacobi functional for general relativity, depending on the radiative data specified at null infinity. In \cref{sec:celestial}, when the flux term \(\Theta \) further can be written as \(p\delta q + \delta k\), we will define the on-shell action as \(S_\mc N -k\).

Under appropriate conditions---namely, for vacuum solutions of Einstein's equations with vanishing cosmological constant in asymptotically flat spacetimes---the bulk Einstein--Hilbert term vanishes on-shell. In this case, the action reduces to boundary contributions, whose precise form depends on the choice of boundary and boundary conditions.

This perspective is closely related to the approach developed by Fabbrichesi, Veneziano, and Vilkovisky \cite{Fabbrichesi:1993kz}, building on earlier work by Arefeva, Faddeev, and Slavnov (AFS) \cite{Arefeva:1974jv}. In that framework, the tree-level $\mathcal{S}$-matrix for gravitational scattering in asymptotically flat spacetimes can be expressed as
\be
\mathcal{S}_{\text{tree}} \sim \exp\!\left(i S\big|_{\text{on-shell}}\right),
\ee
where the action is evaluated on classical solutions satisfying appropriate asymptotic boundary conditions. Under the assumptions stated above, the bulk contribution vanishes and the on-shell action reduces entirely to a boundary functional.
The case of computing boundary terms on null boundaries requires a careful analysis due to the inherent freedom to choose any arbitrary coefficient for the null normal. In the analysis of \cite{Fabbrichesi:1993kz}, this construction was carried out in the absence of gravitational memory, corresponding to configurations with vanishing constant shear mode.  Their action can be written in terms of the Bondi mass aspects of classical gravitational radiation and is given by
\be\label{eq:VV}
S_{v}=-\frac{1}{8\pi G}\left(\int_{\ms{I}^{+}}du\,d^{2}z\,\gamma_{{z}\bar{z}}\, m_{B}-\int_{\ms{I}^{-}}dv\,d^{2}z\,\gamma_{{z}\bar{z}} \,m_{B}\right) 
\ee
With this on-shell action, they reproduced the tree-level four-point amplitude expression in the eikonal approximation. We refer to \Cref{eq:VV} as the hard action for later use. We can simplify the hard action and write it in terms of the Bondi news tensor. Integrating \Cref{eq:massaspectpositive,eq:massaspectnegative} to write the respective mass aspects as
\be
m_B(u)=-\frac{1}{4}\int_{u}^{\infty} du'\,D_A D_B N^{AB}+\frac{1}{8}\int_{u}^{\infty} du'\,N_{AB}N^{AB}
\ee
and
\be
m_B(v)=\frac{1}{4}\int_{-\infty}^{v} dv'\,D_A D_B N^{AB}+\frac{1}{8}\int_{-\infty}^{v} dv'\,N_{AB}N^{AB}
\ee
since we are working with pure gravity, there is no other contribution to the stress tensor. Moreover, when the constant mode is turned off, there will be no contribution to the action from soft radiation. The boundary action now looks like
\be
S_v=-\frac{1}{64\pi G}\int d^2z \gamma_{z\bar{z}}\left[\int_{-\infty}^{\infty} du \int_{u}^{\infty} du' \,N_{AB}N^{AB}-\int_{-\infty}^{\infty} dv \int_{-\infty}^{v} dv' \,N_{AB}N^{AB}\right]
\ee
Integrating by parts
\be
S_v &=-\frac{1}{64\pi G}\int d^2z \gamma_{z\bar{z}}\left[\int_{-\infty}^{\infty} du \left(-N_{AB}C^{AB}-\int_{-\infty}^{u'} du'\,\de_{u'}N_{AB}\,C^{AB}\right)\right.\\
&\qquad - \left.\int_{-\infty}^{\infty} dv \left(N_{AB}C^{AB}-\int_{v}^{\infty} dv'\,\de_{v'} N_{AB}\,C^{AB}\right)\right]
\ee
which simplifies to
\be
S_v = \frac{1}{64\pi G}\int d^2z\gamma_{z\bar{z}}\left(\int_{-\infty}^{\infty} du\int_{u}^{\infty}du'\,\de_{u'}N_{AB}\,C^{AB}-\int_{-\infty}^{\infty} dv\int_{-\infty}^{v}dv'\,\de_{v'}N_{AB}\,C^{AB}\right)
\ee
The boundary action is trivial for an impulsive wave with $C_{AB}$ given by a theta function. Suppose two or more energetic particles (modeled as shockwaves) enter spacetime at time $v=0$ sourced by a stress tensor given by a linear combination of delta functions \footnote{but at different angles. Note that the angle of exit is antipodally matched to the angle of arrival.}
\be
T_{vv}=E_1\delta (v)+E_2\delta (v)
\ee
The corresponding Bondi mass is
\be
m_B(v)=E_1\theta(v)+E_2\theta(v)
\ee
 The geodesics get shifted and the outgoing shockwaves get time delayed showing up at $\ms I^+$ at time $u=u_1$ and $u=u_2$. The Bondi mass at $\ms I^+$ is
 \be
m_B(u)=E_1 \theta (u_1-u)+E_2\theta (u_2-u)
 \ee
 The sign of theta function comes with an extra minus to respect the matching at $i^0$ and the times of arrival $u_1$ and $u_2$ depend on the energies of the particles and the impact parameter. The time delay leads to a non-vanishing $S_v$ and the corresponding amplitude precisely matches the answer known from other methods (see \cite{Fabbrichesi:1993kz} for details).
 
The central goal of this paper is to compute the gravitational boundary action at null infinity in the presence of non-trivial memory effect and to understand its implications for soft theorems. We show that the invariance of the on-shell action under supertranslations leads to the leading Weinberg soft graviton theorem. The inclusion of memory necessitates a careful treatment of corner terms, which are intrinsically linked to changes in the vacuum configuration. In addition, we demonstrate that the flux term plays a crucial role in extending this framework to soft theorems beyond leading order\footnote{We work with the standard IR divergent $\mc{S}$ matrix in this paper.}.
\subsection{Null boundaries}\label{sec:LMPS} A proper treatment of counterterms at null boundaries received little attention over time and was examined in \cite{Neiman:2012fx,Parattu:2015gga} and later by LMPS \cite{Lehner_2016} which we now review. Consider a codimension-1 null hypersurface \(\mc N\) in the physical spacetime defined by $\Phi(x^a) \defn 0$ for some scalar function $\Phi$ that increases towards the future. The null normal to the hypersurface is
\be
\hat n_{a}=-\mu \hat \nabla_{a}\Phi
\ee
 We impose that $\mu > 0$ to ensure that the tangent vector field $ \hat n^{a}$ to the null generators is future pointing. The null generators satisfy the geodesic equation
\be
\hat n^{a}\,\hat \nabla_{a}\hat n^{b}= \hat \kappa\, \hat n^{b}
\ee
where $\hat \kappa$ is the inaffinity which measures the inability of the parameter $\lambda$ along the null generators to be affine. Let $ \hat q_{AB}$ be the induced degenerate metric and $(z,\bar{z})$ be the coordinates on the spatial cross-section of the null segment such that they are constant along the null generators.
Parattu et al\cite{Parattu:2015gga} proposed a boundary action for the null segment of the form
\be\label{eq:parattu}
-2\int_{\mc N} d\hat \Sigma\, (\hat \theta+ \hat \kappa)
\ee
where $\hat \theta$ is the expansion of the null generators and \(d \hat \Sigma\) is the volume element on \(\mc N\) as induced from the physical volume 4-form . Subsequently, LMPS \cite{Lehner_2016} investigated the junctions, or corners, where the null boundary intersects with spacelike or timelike pieces of the boundary of spacetime. They found that at these junctions, due to the lack of smoothness of the normal 1-form, the boundary action acquires additional contributions of the form
\be
\int_{\mc{B}} \alpha \,  \sqrt{\hat q} d^2 z
\ee
with $\alpha=\ln \mu$. Consider a null segment $\mc N$, bounded by a two-surface $\mc{B}_1$ in the past and $\mc{B}_2$ in the future. We fix the induced metric $\hat q_{AB}$ and the null tangent $\hat n^a$ on the boundary segment as part of the boundary conditions\footnote{The boundary conditions that we are working with are equivalent to fixing the intrinsic structure of the null boundary.}. The total variation piece of the boundary action is\footnote{The convention in the section is such that $16\pi G=1$. We will reinstate this scale in the next section.}
\be\label{eq:Sorkinaa}
S_{\mc N}= -2\int_\mc N d\hat \Sigma\, (\hat \theta + \hat \kappa) + \int_{\mc{B}_2} \alpha\, \sqrt{\hat q}\, d^2z - \int_{\mc{B}_1} \alpha\, \sqrt{\hat q} \,d^2z
\ee
The last two terms can be combined to write the action as
\be\label{eq:Sorkinaa2}
S_{\mc N}= -2\int_\mc N d\hat \Sigma\, (\hat \theta + \hat \kappa) + \int d\hat \Sigma\,\Lie_{\hat n} \,\alpha
\ee
It is important to note that this procedure singles out a flux term
\be\label{eq:flux}
\Theta= \hat B_{AB}\delta \hat g^{AB}
\ee
 Here $\hat B_{AB} \defn h_A^a h_B^b \hat B_{ab}$ is the deformation tensor pulled back to cross-sections and is the null hypersurface analog of the extrinsic curvature. At null infinity in asymptotically flat spacetimes, this term becomes the spin-2 pair showing up in the analysis of Ashtekar-Streuebel or the Wald-Zoupas(WZ) flux\cite{Ashtekar:1987tt,Compere:2018ylh}.

 The ambiguity in the boundary action $S_{\mc N}$ comes from the freedom of changing the parametrization of generators $\lambda \rightarrow \lambda'(\lambda,z,\bar{z})$ as well as by the freedom in redefining the function $\Phi \to \Phi'$ such that $\Phi'$ vanishes when $\Phi$ vanishes. Since the coefficient $\mu$ for the null normal is arbitrary in the null case, it can change independently depending upon whether the parameter is changed or the function $\Phi$ is redefined. It turns out that this ambiguity only affects the corner term.
\section{Boundary terms at null infinity}\label{sec:conformal frame}
In this section, we examine the ambiguity in the boundary action at the past and future null infinity of asymptotically flat spacetimes. We show how these ambiguities can be fixed to formulate a boundary action consistent with the memory effect and Weinberg soft theorem. We will work within the conformal spacetime picture of Penrose et al\cite{Penrose:1964ge,Geroch:1977big,Ashtekar:1981bq,Ashtekar:2014zsa,Grant:2021sxk} where null infinity really exists as a codimension-1 null hypersurface. The strategy is to compute the boundary action in the unphysical picture by transforming physical spacetime quantities like the inaffinity and expansion to unphysical ones using the conformal relations. We start with a lightning review of asymptotic flatness and introduce Bondi conformal coordinates following the discussion in \cite{Grant:2021sxk}.

\begin{definition}[Asymptotic flatness at $\ms{I}$] \label{def:asym-flat}
A \emph{physical} spacetime $(\hat M,\hat g_{ab})$, is asymptotically-flat at null infinity if there exists an \emph{unphysical} spacetime $(M,g_{ab})$ with a boundary $\scri = \partial M$ and a diffeomorphism from $M$ to \( \hat M-\ms{I}\) ,
such that
\begin{enumerate}
\item There exists  a smooth function $\Omega$ on $ M$ such that $\ms I$ is defined as \(\Omega\hateq 0\) and \(  \nabla_a\Omega \not\hateq 0\) such that \(g_{ab} = \Omega^2  \hat g_{ab} \) is smooth on \(M\) including at \(\scri\).
\item $\scri$ is topologically $\bb R \times \bb S^2$.
\item The physical metric $\hat g_{ab}$ is the solution to $\Lambda=0$ vacuum Einstein equations $\hat G_{ab}=0$.
\end{enumerate}
\end{definition}
From the vacuum Einstein equations and smoothness of $\Omega$ it follows that $ n_a  n^a \hateq 0$ indicating that $\ms{I}$ is a null hypersurface with a null normal given by $ n_a =  \nabla_a \Omega$ and the vector field $n^a=g^{ab}\, n_b$ serves as the generator for null geodesic on $\ms{I}$ \footnote{Note that there is no minus sign in \(n_a\) because \(\Omega =0\) on \(\ms I^+\) and it increases towards the past.}. It is important to note that the choice of the conformal factor is not unique, we can define $\Omega' \defn \omega\Omega$ with $\omega\vert_{\ms{I}}>0$ leading to another unphysical metric $ g'_{ab}=\omega^2  g_{ab}$. All relevant physical quantities like the BMS charges, and the news tensor which characterizes the gravitational raditation must be independent of the conformal factor. Additionally, through this conformal freedom, we can always choose a divergence free ``conformal frame'' such that  $ \nabla_a  n^a \hateq 0$. From the asymptotic Einstein equations it can be shown that this condition implies $ \nabla_a  n_b \hateq 0$ known as the Bondi condition, which further yields $ n_a  n^a =O(\Omega^2)$. Having
imposed the Bondi condition, the remaining freedom in the conformal factor is of the form
$\Omega \rightarrow \omega\,\Omega$ such that
\be
\omega\vert_{\ms{I}}>0, \qquad \Lie_{ n} \omega\hateq 0
\ee
It is well known that we can geometrically define a coordinate system $( u,\Omega,x^A)$ on $\ms I^+$ for any asymptotically flat spacetime\cite{Ashtekar:1987tt}. The conformal factor $\Omega$ here is chosen such that it satisfies the Bondi condition and makes the metric on the cross-section of $\ms I$ to be the unit round metric on $ \bb S^2$, thus fixing the so-called Bondi frame by using the residual conformal freedom. The line element of the unphysical metric at $\mathscr{I}^+$ in this coordinate system is
\be
d  s^2 \hateq 2d  u\,d\Omega+q_{AB} \,dx^A\, dx^B
\ee
  Instead of fixing the coefficient to unity we express the null normal as
\be
 n_{a}=\mu  \nabla_{a}\Omega
=\mu(0,1,0,0) , \qquad \mu \hateq 1
\ee
Here we have a foliation of the null infinity by the level curves of the function $ u$ which we will take as the retarded time coordinate. We further define the auxiliary normal $ l_{a} \hateq - \nabla_{a}  u$ normal to each cross section such that $ l^{a} l_{a}\hateq 0$ and $ n_{a} \, l^{a}= -1$ to match the conventions of LMPS. \footnote{In fact, the auxiliary normal is a null normal to the $ u=\text{const.}$ hypersurfaces away from $\ms I^+$ as well. Moreover LMPS normalization is in physical spacetime, i.e., they have \(\hat n_a \, \hat l^a =-1\). But here we have chosen the normalization of \(n_a\) and \(l^a\) such that \(\hat n_a\, \hat l^a = -1= \hat n_a \, \hat l^a\).} Now we can extend this coordinate system to the neighborhood of $\mathscr{I}^+$ by Lie dragging the coordinates $( u,x^A)$ along the transverse vector field $ l^{a}$ and fixing the Bondi gauge on $( u,\Omega,x^A)$\cite{Grant:2021sxk}. By definition, $ u$ is an affine parameter on $\mathscr{I}^+$, 
  \be
   n^{a} \nabla_{a}  u\hateq 1
  \ee
  Now if we set $\mu$ to unity we won't have any corner terms in our action. Instead we work with $\mu=\frac{1}{ g^{\Omega u}}$. This choice of $\mu$ is well motivated because the foliation of $\ms{I^+}$ with the level sets of $ u$ provides this choice of normalization\cite{Oliveri:2019gvm}\footnote{Obviously, the foliation of $\ms I^+$ is not unique but this choice of foliation and of $\mu$ is the standard choice.}. In the Bondi gauge, $\Omega$ is a ``radial'' coordinate. This means that the coefficient $\mu$ will be metric-dependent. Alternatively, we can also work in the Newman-Unti gauge\cite{Newman:1962cia} where $\Omega$ is an affine parameter along the transverse null geodesics. In the Newman-Unti gauge $\mu=1$, see \Cref{sec:NewmanUnti} for a discussion on boundary terms.
  
Imposing the Bondi condition $ \nabla_{a}  n_{b}\hateq 0$ on $\mathscr{I^+}$ by using the freedom in conformal scaling implies $  n_a  n^a=O(\Omega^2)$. It can be easily seen that this amounts to 
\be
 g_{uu}=O(\Omega^2), \qquad  g^{\Omega\Omega}=O(\Omega^2)
\ee
Working with $\mu= g_{u\Omega}$, we have $ n^a \hateq \partial_u$, and
$ n_a  n^a= g^{\Omega \Omega}$. The geodesic equation for the null geodesics is
\be
 n^{a} \nabla_{a} n^{b}= \kappa  n^{b}, \qquad
\ee
$ \kappa \hateq 0$ since the null generators are affine parametrized by construction. A general and covariant-looking expression for the inaffinity $ \kappa$ is\cite{Parattu:2015gga}
\be\label{eq:kappa}
 \kappa=\Lie_{ n} \ln\mu-\frac{1}{2}\Lie_{ l}  n_a  n^a
\ee
and the corner term defined in the LMPS action is
\be\label{eq:alpha}
\alpha=\ln \mu
\ee
 Note that $ n_a  n^a\hateq 0$ but it is nonzero off the surface and since $\mu$ is metric dependent, the inaffinity $ \kappa$ will be non-zero at $O(\Omega)$. The strategy is to compute boundary action on a \(\Omega = \text{const.}\) null hypersurface in physical spacetime and push it to null infinity \footnote{To take $O(\Omega)$ corrections into account, we can think of the spacetime being foliated by a family of timelike hypersurfaces in the vicinity of $\ms I$ (which is $\Omega \hateq 0$), with normal $  n_a =\mu\, \nabla_a \Omega$ to each surface and take $\Omega \to 0$ limit at the end. Parattu et al\cite{Parattu:2015gga} have shown that this kind of limiting process gives the same boundary action.\label{omegacorrections}}. Using \Cref{eq:Sorkinaa2}, the boundary action on a null hypersurface \(\mc N\) in physical spacetime is
\be
S_\mc N=\frac{1}{16\pi G}\int_{\mc N} d\hat \Sigma\,(-2(\hat \theta + \hat \kappa) +\Lie_{\hat n}\alpha)
\ee
where $d\hat \Sigma$ is the volume 3-form on \(\mc N\). 
The inaffinity is ambiguous even for a fixed conformal factor as it depends upon the choice of a normal representative via its dependence on $\mu$. Now we will compute these quantities in the unphysical picture and relate them to physical ones to write down the boundary action.
The most general form of the unphysical metric in conformal Bondi-Sachs coordinates $( u,\Omega,x^A)$ is given by\cite{Grant:2021sxk,Madler:2016xju} 
\be\label{eq:unphys-BS}
	d s^2 \equiv - \Omega^3 V e^{2 \beta} d u^2 + 2 e^{2 \beta} d\Omega\, d u +  h_{AB} (dx^A - U^A d u )(dx^B - U^B d u )\,,
\ee
We consider the following expansion of the metric components
\be\label{eq:bondi-expansion}
\Omega^3 V = \Omega^2 - 2\Omega^3 m_B+ O(\Omega^4) \eqsp& U^A =  - \frac{\Omega^2}{2}  D_BC^{AB}  + O(\Omega^3)\,, \\
\beta =   - \frac{\Omega^2}{32} C_{AB} C^{AB} + O(\Omega^3) \eqsp&  h_{AB} = q_{AB} + \Omega C_{AB} +  \frac{\Omega^2}{4}q_{AB}C_{CD} C^{CD} + O(\Omega^3)\,.
    \ee
    The Weyl Scalar \(\Omega^{-1}C_{nlnl}\) is
    \be \label{eq:weylscalar}
    \mc P \hateq -2m_B + \frac{1}{4}N_{AB}C^{AB}, \qquad \text{where} \qquad N_{AB} = -\partial_u C_{AB}.
    \ee
    Furthermore using $
 l^a= g^{ab} l_b$, we have $ l^{\Omega}=- g^{u\Omega}$ and from the metric expansion $\mu=e^{2\beta}$. The expansion of generators is
\be
\theta \defn \frac{1}{2} h^{AB}\Lie_{ n}  h_{AB} = D_A U^A
\ee
The inaffinity and the corner term are
\be
\kappa
= \Omega + \Omega^2\left(\frac{1}{8} N_{AB}C^{AB} - 3 m_B\right) + O(\Omega^3).
\ee
\be
\Lie_{ n}\alpha = \frac{\Omega^2}{8} N_{AB}C^{AB} + O(\Omega^3)
\ee
Using the conformal relations between physical and unphysical quantities \cite{Prabhu:2026kps}
\be
\hat \kappa &= \Omega^2  \kappa + \Omega  n_a  n^a, \qquad \Lie_{\hat n}\alpha = \Omega^2 \Lie_{ n} \alpha \\
\hat \theta &= \Omega^2  \theta - 2\Omega  n_a  n^a 
\ee
so that the boundary action integrand is 
\be
\Omega^4 (2m_B + D_A D_B C^{AB} - \frac{1}{8}N_{AB}C^{AB}) + O(\Omega^5)
\ee
Interestingly, we do not get a radially divergent term proportional to the Euler number of the sphere metric as obtained by Compere et al \cite{Compere:2018ylh}. Anyways that term would have been a Minkowski contribution and we could have used the reference subtraction scheme commonly used in covariant phase space formalism. In the covariant phase space formalism, charges and fluxes associated with asymptotic symmetries are defined relative to a reference spacetime. In the present context, the reference spacetime is taken to be any conformal completion of Minkowski spacetime.
Using the conformal relation between physical and unphysical volume 3-forms, \(d \hat \Sigma = e^{2\beta}\Omega^{-4}d\Sigma\), and substituting this in the boundary action \Cref{eq:Sorkinaa}, we get finite boundary term \footnote{The \(D_A D_B C^{AB}\) term vanishes when integrated on the sphere.}
\be
S_k\rvert_{\ms I^+} \hateq \frac{1}{16\pi G}\int_{\ms I^+} du\,d^2z\,\gamma_{z\bar{z}}\,\left(2m_B -\frac{1}{8}N_{AB}C^{AB}\right)
\ee
 A similar procedure can be followed at $\mathscr{I}^-$ to get the boundary term
\be
S_k\rvert_{\ms I^-} \hateq \frac{1}{16\pi G}\int dv\, d^2z \,\gamma_{z\bar{z}} \,\left(2m_B+\frac{1}{8}N_{AB}C^{AB}\right)
\ee
Following LMPS convention of combining the boundary terms, the total boundary action is\footnote{The integrable piece $S_k$ is sometimes called the boundary Lagrangian. The LMPS convention of combining boundary terms is such that the total action is additive, i.e., suppose that we have a manifold $\mc{M}$ divided into two pieces $\mc{M}_1$ and $\mc{M}_2$, then the partition function $Z[\mc{M}]$ should be equal to the product of partition functions $Z[\mc{M}_1]Z[\mc{M}_2]$.}
\be
S_{k}& \hateq \frac{1}{8\pi G}\left(\int_{\mathscr{I}^{+}}du\,d^{2}z\,\gamma_{z\bar{z}}\, m_{B}+\int_{\mathscr{I}^{-}}dv\,d^{2}z\,\gamma_{z\bar{z}}\, m_{B}\right)\\ 
&  +\frac{1}{128\pi G}\left(\int d^{2}z\,\gamma_{z\bar{z}}\,\overset{0^{out}}{N_{AB}}\,\hat{C}^{AB}-\int d^{2}z\,\gamma_{z\bar{z}}\,\overset{0^{in}}{N_{AB}}\,\hat{C}^{AB}\right) \,
\ee
where we have written the last line as corner terms using the definitions in \cref{sec:prelim}, see the discussion around \cref{eq:shear}.
The flux term \Cref{eq:flux} in Bondi coordinates is\footnote{We also get $\frac{\mc{R}}{4}q_{AB}\delta C^{AB}$, but it vanishes using $p\delta q=-q\delta p$ and the boundary conditions. This will contribute to the symplectic potential when we work with generalized BMS (gBMS).}
\be\label{eq:flux}
\Theta \hateq -\frac{1}{32\pi G} N_{AB}\,\delta C^{AB}
\ee
For spacetimes without radiation, the flux vanishes. On the other hand, we cannot restrict ourselves to the spacetimes where the variation of $C_{AB}$ vanishes since the shear sources everything in pure gravity. Moreover, interesting boundary degrees of freedom arise when we allow $C_{AB}$ to vary as shown by Strominger\cite{Strominger:2013jfa}. Under supertranslations parametrized by a function $f(z,\bar{z})$ on the sphere, the spacetime undergoes a vacuum transition with the constant shear mode transforming as $\mc{C} \to \mc{C}+f$ while adding a soft graviton, the latter being viewed as a Goldstone boson accompanying the breaking of supertranslation symmetry. Consequently, the counter terms in the presence of radiation have this non-integrable piece and a good variational principle in the usual sense may not exist.

If we restrict ourselves to spacetimes with a trivial memory mode, we recover the `hard action' derived by Fabbrichesi et al \cite{Fabbrichesi:1993kz}, albeit with a sign difference\footnote{This case is trivial in the sense that even in classical gravitational scattering, memory effect exists. The sign difference comes from the use of the Gauss theorem. It is just a matter of sign convention, in our computations we demanded that the tangent vectors be future pointing for both past and future null infinity.}.  
\subsection{Corner term ambiguity}\label{sec:corner ambiguities}
As noted by LMPS, there are two types of ambiguities in the boundary action at null boundaries: reparametrization freedom from $\lambda$ to $\lambda'(\lambda,x^A)$ and changing the embedding from $\Phi$ to $\Phi'$. Since we are working in the conformal picture at $\mathscr{I}^+$, we will keep the affine parameter $u$ intact. On the other hand, while we appear to have exhausted the conformal freedom in redefining the conformal factor to fix the Bondi frame, there remains residual freedom associated with these transformations. Consider a change in the (Bondi fixed) conformal factor, $\Omega \rightarrow \Omega'\defn\gamma\,\Omega$ such that $\gamma \hateq1$ and $\Lie_n \gamma \hateq 0$. An explicit example of such a factor is $\gamma=e^{m\beta}$ where $m \in \bb R$ and $\beta$ is given by \Cref{eq:bondi-expansion}. The metric $g_{ab}\rvert_{\ms I}$ is unchanged and $\Omega' \hateq 0$ when $\Omega \hateq 0$. In this case, $\mathscr{I}^+$ will be defined by $\Omega' \hateq 0$. The null normal can be re-expressed as
\be
n_{a} =\mu'\nabla_{a}\Omega'
\ee
where $\mu'$ is related to the old $\mu$ by \be\mu'=\mu\frac{d\Omega}{d\Omega'}
\ee
Define $A \defn \frac{d\Omega}{d\Omega'} $, then we have
\be\label{eq:A}
A^{-1} = \frac{d\Omega'}{d\Omega}=\gamma+\Omega\frac{d\gamma}{d\Omega}
\ee
The ``corner'' term $\alpha$ and inaffinity transform as
\be
\alpha &\rightarrow \alpha' =\alpha+\ln A \\
\kappa &\rightarrow \kappa' =\kappa + \Lie_n \ln A
\ee
Since $\gamma \hateq 1$, the boundary metric $\hat q_{AB}$ and its determinant $\sqrt{\hat q}$ are unchanged.  
The boundary action in this new parametrization is
\be
S_{\mc N}=\frac{1}{16\pi G}\int d\hat \Sigma\, \left(-2(\hat \theta' + \hat \kappa') + \Lie_{\hat n}\alpha'\right)
\ee
As can be checked that the hard action \Cref{eq:VV} does not change for the class of $\gamma$ we have. Considering that, we will have a ``class'' of on-shell actions with a variable corner term due to corner ambiguities encoded in \(A\). This implies that the boundary action will look like
\be
S_{k}& \hateq \frac{1}{8\pi G}\left(\int_{\mathscr{I}^{+}}du\,d^{2}z \,\gamma_{z\bar{z}}\,m_{B}+\int_{\mathscr{I}^{-}}dv\,d^{2}z\,\gamma_{z\bar{z}}\, m_{B}\right)\\ 
& +\frac{\zeta}{128\pi G}\left(\int d^{2}z\,\gamma_{z\bar{z}}\,\overset{0^{out}}{N_{AB}}\,\hat{C}^{AB}-\int d^{2}z\,\gamma_{z\bar{z}}\,\overset{0^{in}}{N_{AB}}\,\hat{C}^{AB}\right) \,
\ee
where $\zeta \in \bb R$. We want to fix $\zeta$ by demanding that the on-shell action reproduces the 5-point amplitude with a soft graviton insertion. Writing the tree-level $\mc S$ matrix as
 \be
\mc{S}_{tr}=e^{iS_k}= e^{iS_v+iS_{\text{soft}}}
 \ee
where $S_v$ is the non-zero hard term for $2-2$ massless scattering , \Cref{eq:VV} and the soft part of the action is
 \be
S_{\text{soft}}= \frac{\zeta}{128\pi G}\left(\int d^{2}z\,\gamma_{z\bar{z}}\,\overset{0^{out}}{N_{AB}}\,\hat{C}^{AB}-\int d^{2}z\,\gamma_{z\bar{z}}\,\overset{0^{in}}{N_{AB}}\,\hat{C}^{AB}\right) \,
 \ee
As part of the initial data, we can set the memory of the incoming state to be zero, i.e., $\overset{0^{in}}{N_{AB}}=0$. The symplectic form on the extended Ashtekar-Streubel phase space consistent with the leading soft theorem is (see \cref{sec:CFS})
\begin{equation}\label{eq:Symplecticform}
\Omega_{\text{CS}}
=
-\frac{1}{32\pi G}
\left(\int_{\mathscr I^+} du\, d^2z \,\gamma_{z\bar z} \;
\delta N^{AB} \wedge \delta C_{AB}
+
\int_{S^2} d^2z \, \gamma_{z\bar z} \,
\delta(\overset{0}{N}_{AB}) \wedge \delta(\hat C^{AB})\right)
\end{equation}
 The constant shear $\hat{C}_{AB}\rvert_{\ms I^+}=- 2D_A D_B\mc{C}$ and leading memory $\overset{0}{N}_{AB}= -\int du \, N_{AB}$ is a conjugate pair on the `soft' phase space. In the generating functional approach of the $\mc{S}$ matrix, a functional derivative with respect to $\hat{C}_{AB}$ will create a soft graviton insertion\cite{Arefeva:1974jv,Kim:2023qbl}
\be
i\frac{\delta}{\delta \pi_{\mc{C}}}\mc{S}_{tr}[\overset{0}{N}, \mc{C}]=\frac{\zeta\,\gamma_{z\bar{z}}}{8G}\int_{-\infty}^{\infty} du\,\langle \text{out}|\left(N^{\bar{z}\bar{z}}\hat{\mc{S}}_{tr}-\hat{\mc{S}}_{tr}N^{\bar{z}\bar{z}
}\right)|\text{in}\rangle
\ee
where $\pi_{\mc{C}}=\frac{1}{32\pi}\hat{C}_{AB}$ and $\hat{\mc{S}}_{tr}$ is the corresponding operator. In \cite{Choi:2017ylo}, it was shown that the supertranslation invariance of the $\mc{S}$ matrix gives(in our convention of the news tensor)
\be
-\frac{\gamma_{z\bar{z}}}{4G}\int_{-\infty}^{\infty} du\,\langle \text{out}|\left(N^{\bar{z}\bar{z}}\hat{\mc{S}}_{tr}-\hat{\mc{S}}_{tr}N^{\bar{z}\bar{z}
}\right)|\text{in}\rangle= -\sum_{i} \eta_i \frac{p_i^a p_i^b}{i(k\cdot p_i-i\epsilon)}\langle \text{out}\lvert \hat{\mc{S}}_{tr}\rvert \text{in}\rangle
\ee
where $\langle \text{out}\lvert \hat{\mc{S}}_{tr}\rvert \text{in}\rangle=\mc{M}(p_1,p_2)$ is the four-point amplitude with the leading soft factor factorized out. Following Choi-Akhoury, we have restricted to a single helicity sector and retained only the \(\bar z \bar z\) components, taking the two helicity components contribution into account equally. This fixes $\zeta=-2$ and the on-shell action is\footnote{The mass aspect $m_B$ is not a phase space variable, but it can be written in terms of news which is.}
\be\label{eq:onshell}
S_{k}& \hateq \frac{1}{8\pi G}\left(\int_{\mathscr{I}^{+}}du\,d^{2}z\,\gamma_{z\bar{z}}\, m_{B}+\int_{\mathscr{I}^{-}}dv\,d^{2}z\,\gamma_{z\bar{z}}\, m_{B}\right)\\ 
& - \frac{1}{64\pi G}\left(\int d^{2}z\,\gamma_{z\bar{z}}\,\overset{0^{out}}{N_{AB}}\,\hat{C}^{AB}-\int d^{2}z\,\gamma_{z\bar{z}}\,\overset{0^{in}}{N_{AB}}\,\hat{C}^{AB}\right) \,
\ee 
Another check we can perform is to see whether the action \Cref{eq:onshell} is supertranslation invariant. Using the variations of the Bondi masses \Cref{eq:mbfuture}, \Cref{eq:mbpast} and those of $C_{AB}$ \Cref{eq:futureshear}, \Cref{eq:pastshear}, the fact that memory $\overset{0}{N}_{AB}$ does not change under supertranslations and matching conditions across spatial infinity, we can compute the variation of supertranslation invariance of the combined boundary action
\be\label{eq:supertranslationinvariance}
\delta_f S_k &\hateq \frac{1}{8\pi G}\left(\int_{\ms{I}^{+}}du\,d^{2}z\,\gamma_{{z}\bar{z}}\,f\de_u m_{B}+\int_{\ms{I}^{-}}dv\,d^{2}z\,\gamma_{{z}\bar{z}}\, f\de_v m_{B}\right)\\ 
&+\frac{1}{32\pi G}\left[\int d^{2}z\,\gamma_{{z}\bar{z}}\left(\overset{0^{out}}{N^{AB}}D_{A}D_{B}f +2D_{A}fD_{B}\overset{0^{out}}{N^{AB}}\right)\right.\\
&\left.+\int d^{2}\,z\gamma_{{z}\bar{z}}\left(\overset{0^{in}}{N^{AB}}D_{A}D_{B}f+2D_{A}fD_{B}\overset{0^{in}}{N^{AB}}\right)\right] \\ 
&-\frac{1}{64 \pi G}\left[-2\int d^{2}z\,\gamma_{{z}\bar{z}}\,\overset{0^{out}}{N^{AB}}D_{A}D_{B}f-2\int d^{2}z\,\gamma_{{z}\bar{z}}\,\overset{0^{in}}{N^{AB}}D_{A}D_{B}f\right]=0 \,
\ee
Thus
\be
\delta_f S_k =0 , \qquad \text{Or} \qquad [Q_f, S_k]=0
\ee
where $Q_f$ generates supertranslations on the radiative phase space. Interestingly, the boundary action's supertranslation invariance looks like the supertranslation charge's\footnote{By supertranslation charge, we mean the integrable part of it. Supermomentum has a nonintegrable part which is related to the radiative flux leaking through the boundary and can be removed locally and covariantly using Wald Zoupas prescription.} conservation up to a factor of 2. Moreover, this implies in the quantum theory that
\be
\langle \text{out}\lvert\left( Q_f^+\hat{\mc{S}}_{tr}-\hat{\mc{S}}_{tr}Q_f^-\right)\rvert\text{in}\rangle=0
\ee
Strominger conjectured that the BMS group is an exact symmetry of the full quantum gravitational $\mathcal{S}$-matrix~\cite{Strominger:2013jfa}. In the present analysis, we have shown explicitly that supertranslations act as symmetries of the tree-level $\mathcal{S}$-matrix. This follows from the fact that the gravitational memory effect and the Weinberg soft graviton theorem exist already at tree level and do not receive loop corrections. $\zeta=-2$ implies that
\be
A= e^{-2\beta}, \qquad \gamma=e^{-\frac{2}{3}\beta}
\ee
 and the correct corner term is
\be
\Lie_n \alpha' = \frac{\Omega^2}{4}N_{AB} C^{AB} + O(\Omega^3)
\ee
\section{Subleading symmetry}\label{sec:celestial}
In this section, we consider the extension of the BMS group to include (holomorphic) superrotations as the asymptotic symmetries on $\ms I^+$. In this scenario, the superrotations are generated by holomorphic vector fields that satisfy the conformal killing equation on the sphere.  Unlike meromorphic superrotations (the full extended BMS group, eBMS), with a finite number of isolated poles, the holomorphic superrotations have a singularity only at the north pole of the sphere.  The holomorphic vector fields leave the boundary metric $q_{AB}$ invariant but act non trivially on the space of (STF) Geroch tensors that we define below. Therefore, the superrotation vector fields are best studied in an arbitrary conformal frame. Consider time-independent conformal transformations of the form $\gamma_{AB} \to q_{AB} \defn e^{2\omega(x^A)}\gamma_{AB}$ where $q_{AB}$ is not constrained to have the Ricci scalar fixed to $\mc{R}=2$. In that case,
\be
{g}_{uu}&=\frac{\mc{R}\Omega^2}{2}- 2m_{B} \Omega^3+O(\Omega^4)
\ee
and the $O(\Omega)$ correction to the boundary metric $q_{AB}$ grows a linear in $u$ component\cite{Compere:2018ylh,Campiglia:2021bap}
\be
C_{AB}(u,z,\bar{z})=\sigma_{AB}(u,z,\bar{z})+\hat{C}_{AB}(z,\bar{z})+(u+\mc{C}) T_{AB}(z,\bar{z})
\ee
such that\footnote{For the discussion in this section, we will need more relaxed fall offs for the news, i.e., $N_{AB} \sim O(u^{-2-\epsilon})$ as $|u| \to \infty$.} 
\be\label{eq:integrand}
\int du \,\de_u\left(C_{AB}C^{AB}\right)=-2\overset{0}{N}_{AB}\hat{C}^{AB} - 2\overset{0}{N}_{AB}\,\mc{C}T^{AB}+ 2\int du \,T_{AB}C^{AB}_{\text{vac}}
\ee
where $C^{AB}_{\text{vac}}$ is the vacuum shear
\be
C^{AB}_{\text{vac}}= \hat{C}^{AB}(z,\bar{z})+(u+\mc{C}) T^{AB}(z,\bar{z})
\ee
and $T_{AB}$ is the trace-free part of the symmetric Geroch tensor $\rho_{AB}$ constructed from $q_{AB}$ such that\cite{Geroch:1977big,Campiglia:2014yka,Campiglia:2021bap}
\be\label{eq:gerochconditions}
\rho_{AB}q^{AB}=\mc{R}[q], \qquad D_{[A}\rho_{BC]}=0
\ee
and the conformally invariant news is\cite{Ashtekar:1981bq}
\be
N_{AB}=-\de_u C_{AB}-T_{AB}
\ee
The Geroch Tensor can be split into a trace-free and trace part as
\be
\rho_{AB}=-T_{AB}+\frac{\mc{R}[q]}{2}q_{AB}
\ee 
It is instructive to rewrite the integral, \Cref{eq:integrand} as
\be
\int du \,\de_u\left(C_{AB}C^{AB}\right)=-2\overset{0}{N}_{AB}\hat{C}^{AB} - 2\overset{0}{N}_{AB}\,\mc{C}T^{AB} - 2\overset{1}{N}_{AB}T^{AB}+ 2 \int du\, C_{AB}T^{AB}
\ee
The last term can be integrated to give
\be\label{eq:Gerochint}
2 \int du C_{AB}T^{AB}= 2\overset{1}{N}_{AB}T^{AB}+ 2\int du \,T_{AB}C^{AB}_{\text{vac}}
\ee
so it cancels the subleading soft news term and has an infinite contribution from the second term which is nothing but the mass aspect of the vacua due to the presence of the ``source'' $T_{AB}$ \footnote{Since the vacua have zero energy, the presence of $T_{AB}$ is incompatible with the stability of vacua. We have checked that it also spoils the supertranslation invariance of the boundary action in a non-Bondi frame. Compere and Long\cite{Compere:2016jwb} put forward the Dirichlet boundary condition $T_{AB}=0$ so that the vacuum is stable.\label{ft19}}. So we get our original answer back in \Cref{eq:integrand}. 
However, we have not accounted for an extra contribution from the flux ter $\Theta$, \Cref{eq:flux}. The flux term is not \emph{conformally invariant} since $\de_u C_{AB}$ is not. We can separate the Geroch part $T_{AB}\delta C^{AB}$ to make it conformally invariant. As a result, this term gives us
\be
-\frac{1}{32\pi G}N_{AB}\delta C^{AB} - \frac{1}{32\pi G}\delta (T_{AB} C^{AB}) + \frac{1}{32\pi G}C^{AB}\delta T_{AB}
\ee
The first term is the conformally invariant WZ flux term when we are \emph{not} in the Bondi frame\cite{Odak:2022ndm}. The second term is a complete variation that will contribute to the action. Suppose we stick to globally defined conformally killing vector fields (CKV). In that case, Geroch has proved that $T_{AB}$ is universal, i.e., $\delta T_{AB}=0$\footnote{It is to be noted that $\rho_{AB}$ is anomalous, i.e., its field space transformation under a diffeomorphism differs from the Lie derivative. The anomaly in $T_{AB}$ is exactly equal to that of $-\de_u C_{AB}$ making $N_{AB}$ covariant.}. Hence the third term vanishes if we have the sphere topology (with no punctures). It turns out that after taking the second term into account, the boundary action at $\ms I^+$ looks like
\be
S_k\rvert_{\ms I^+}=\frac{1}{16\pi G}\int_{\ms I^+} du\,d^2z\,\sqrt{q}\,\left(2m_B-\frac{1}{4}N_{AB}C^{AB}+\frac{1}{2} T_{AB}C^{AB}\right)
\ee
Using \Cref{eq:integrand,eq:Gerochint}, the boundary term becomes
\be\label{eq:bageroch}
S_k\rvert_{\ms I^+}=\frac{1}{8\pi G}\int_{\ms I^+} du\,d^2z\,\sqrt{q}\,M   + \frac{1}{64\pi G}\int d^2z\,\sqrt{q}\left(\overset{0^{out}}{N_{AB}}\hat{C}^{AB}+\overset{out}{\Pi}_{AB}T^{AB}\right)
\ee
where we have defined $\Pi_{AB}=2\overset{1}{N}_{AB}+\mc{C}\overset{0}{N}_{AB}$ and $M=m_B +\frac{1}{8} T_{AB}C^{AB}_{\text{vac}}$.
Taking the contribution coming from $\ms I^-$ into account, the final action is
\be
S_{k}&=\frac{1}{8\pi G}\left(\int_{\mathscr{I}^{+}}du\,d^{2}z\,\sqrt{q}\, M+\int_{\mathscr{I}^{-}}dv\,d^{2}z\,\sqrt{q}\, M\right)\\ 
& + \frac{1}{64\pi G}\left[\int d^{2}z\,\sqrt{q}\,\left(\overset{0^{out}}{N_{AB}}\,\hat{C}^{AB}+\overset{out}{\Pi}_{AB}T^{AB}\right)-\int d^{2}z\,\sqrt{q}\,\left(\overset{0^{in}}{N_{AB}}\,\hat{C}^{AB}+\overset{in}{\Pi}_{AB}T^{AB}\right)\right] \,
\ee
  Campiglia and Laddha\cite{Campiglia:2021bap} gave a construction of the extended phase space by including the superrotation modes. The symplectic form on the extended `soft' phase space (restricted to the Geroch-subleading memory sector)  is
\be
\Omega_{\Gamma_e}(\overset{1}{N_1}, \overset{1}{N_2} )=\frac{1}{32\pi G}\int d^2z \,\sqrt{q}\left(\overset{out}{\Pi}_{1\,AB}T^{AB}-\overset{out}{\Pi}_{2\,AB}T^{AB}\right)
\ee
 Working with the generating functional approach, we assume that the supertranslation conservation law constrains the tree-level $\mc{S}$ matrix, i.e., we have already taken a functional derivative with respect to $\pi_{\mc{C}}$ and set $\mc{C}=0$. Then a functional derivative with respect to $\pi_T=\frac{1}{16\pi}T^{AB}$ of the $\mc{S}$ matrix will give a subleading soft graviton insertion with the correct subleading soft factor\footnote{See Equation 5.3 of \cite{Donnay:2022hkf}. As pointed out in \cite{Campiglia:2021bap}, the quantum operator $\overset{1}{\widehat{N}}_{zz}$ is a sum of the usual zero frequency subleading graviton operator and a soft operator linear in $\frac{\delta}{\delta T_{zz}}$. For holomorphic superrotations, it acts as the latter. For a generic superrotation, only one of these terms contributes to the soft superrotation charge.}.
 
However, it is important to mention that the flux term is not covariant when holomorphic superrotations are included as asymptotic symmetries. As pointed out before, Geroch demonstrated the universality of $\rho_{AB}$ under the crucial assumption of sphere topology. When this assumption is relaxed to allow a punctured sphere topology, $\rho_{AB}$ is neither unique nor universal, implying $\delta \rho_{AB} \neq 0$. The new degrees of
freedom in Geroch’s tensor can be interpreted in terms of a conformal field theory for which $T_{AB}$ is
the stress-energy tensor\cite{Compere:2018ylh}. On the punctured sphere, which can be decompactified to a plane with the singularity at infinity\footnote{Working on the complex plane, we have $\mc{R}=0$ and the components of boundary metric are $q_{z\bar{z}}=1$, and $q_{zz}=0=q_{\bar{z}\bar{z}}$. Moreover $D_A=\de_A$ and $\hat{C}_{AB}=(-2\de_A\de_B+T_{AB})\mc{C}$,  the full term in the brackets is called the superrotation covariant derivative. See \cite{Campiglia:2020qvc,Campiglia:2024uqq} for details.}, $\rho_{AB}=0$ emerges as a trivial solution to \Cref{eq:gerochconditions} (also see \Cref{ft19}). Non-trivial $\rho_{AB} \neq 0$ sectors can be generated from this trivial solution by considering spacetime diffeomorphisms generated by finite superrotations\footnote{In \cite{Campiglia:2021bap}, authors considered the space of Geroch tensors which are holomorphic (and anti-holomorphic) in the entire complex plane. In general, they are meromorphic with poles of order 2. In stereographic coordinates, $T_{zz}$ is the holomorphic one, i.e., $\de_{\bar{z}}T_{zz}=0$.}. Hence the new flux term when (holomorphic) superrotations are allowed is
\be
\Theta=-\frac{1}{32\pi G}N_{AB}\delta C^{AB} + \frac{1}{32\pi G}C^{AB}\delta T_{AB}
\ee
Recently, Rignon-Bret and Speziale\cite{Rignon-Bret:2024gcx} demonstrated that the second term is covariant up to a total divergence on the cross section. Further they showed that, for the trivial choice $\rho_{AB}=0$, we have $\delta_{\xi} \rho_{AB}=0$ in any conformal frame where $\xi$ is a globally defined CKV. Whether such a choice of $\rho_{AB}$ exists for holomorphic (and meromorphic) superrotations is unclear to the author. We encourage readers to see \cite{Rignon-Bret:2024gcx} for technical details.

The situation becomes a little complicated when considering the generalized BMS (gBMS) group. The vector fields generating $\text{Diff}(S^2)$ are not CKV and do not preserve the universal structure at $\ms I$ by altering the 2d boundary metric. So it is important to treat $q_{AB}$ in the gBMS picture as a dynamical variable on the phase space along with the news and shear\footnote{Shear is strictly not a phase space variable since it does not correspond to a local observable in the bulk and its Poisson bracket is not well defined \cite{He:2014bga,Campiglia:2015yka,Prabhu:2022zcr}. However HLMS \cite{He:2014bga} implemented the C-K condition to define a nice Poisson bracket and fix the ambiguity in the phase space transformation of \(C_{AB}\) under supertranslations.} This leads to an extra mode $p_{AB}$ which is conjugated to $q^{AB}$ on the extended phase space, supplementing the usual modes\cite{Campiglia:2020qvc,Campiglia:2024uqq}. We did not obtain such a term in the action.  Since $T_{AB}$ depends upon $q_{AB}$ and $\delta q_{AB} \neq 0$, the issue of the non-covariance of the flux term is present here as well. Note that $\Theta$ will have extra terms along with the usual WZ flux due to the condition $\delta q_{AB} \neq 0$. 
\subsection{$\text{sub}^n$-leading symmetry}\label{sec:subsubleading}
Consider a `generalization' of the (STF) Geroch tensor
\be\label{eq:generalizedGeroch}
T^{AB} \to T_g^{AB} =  \sum_{n=0}^{\infty}u^n \,T_{n+1}^{AB}, \qquad T_1^{AB} \defn T^{AB}
\ee
so that\footnote{for simplicity, we have replaced $u+\mc{C}$ by $u$.} 
\be
C^{AB}=\sigma^{AB}+\hat{C}^{AB}+ \sum_{n=0}^{\infty}u^{n+1}\,T_{n+1}^{AB}
\ee
Recall that the STF tensor $T_{AB}$ was defined by the condition\cite{Campiglia:2024uqq,Freidel:2021ytz}
\be\label{eq:gerochcondition2}
D_A T^{AB} + \frac{1}{2} D^{B}\mc{R}=0
\ee
\Cref{eq:gerochcondition2} is obeyed at each order in $u$ if we impose that $ \{T_n^{AB}; n\geq 2\}$ are divergence-free STF tensors on the sphere\footnote{$\{T_n^{AB}\}$ vanish in the Bondi frame.}. On a compact sphere with smooth tensor fields, a symmetric trace-free tensor satisfying $D_A S^{AB}=0$ vanishes identically. However, in the present setup, we allow punctures on the sphere, as required for holomorphic or meromorphic superrotations. In this case, the tensors $T_n^{AB}$ are not globally smooth and can have singularities (e.g.\ poles) at isolated points. The divergence-free condition is then understood away from the punctures, and non-trivial solutions exist. In stereographic coordinates, these correspond to holomorphic (or meromorphic) components such as $T_{zz}(z)$ with $\partial_{\bar z} T_{zz}=0$. $\rho_{AB}$ is now time-dependent but still defined by the Geroch conditions \Cref{eq:gerochconditions}. In this case, the conformally invariant news is
\be
N^{AB}=-\de_u C^{AB}-\sum_{n=0}^{\infty}(n+1)u^n\,T_{n+1}^{AB}
\ee
We can compute the integral
\be
\int du\, \de_u \left(C_{AB}C^{AB}\right)=-2\overset{0}{N}_{AB}\hat{C}^{AB}+...
\ee
where ... is the integral over $\ms I^+$ of the generalized mass aspect of the vacua (see \Cref{eq:integrand}). On the other hand, the covariantization of flux gives us
\be
\Theta= -\frac{1}{32\pi G}N_{AB}\delta C^{AB}+\frac{1}{32\pi G}C^{AB}\left(\sum_{n=0}^{\infty}(n+1)u^n \delta T_{n+1,AB}\right)
\ee
And the extra contribution to the boundary action is
\be\label{eq:extendedboundaryterms}\frac{1}{32\pi G}\int du\, C_{AB}\left(\sum_{n=0}^{\infty}(n+1)u^{n}\,T_{n+1}^{AB}\right)= \frac{1}{32\pi G}\left(\overset{1}{N}_{AB} T_1^{AB}+\overset{2}{N}_{AB}T_2^{AB}+\overset{3}{N}_{AB}T_3^{AB}+...\right)
\ee
where $...$ contains next to $\text{sub}^3$-leading terms as well as the generalized mass aspect of the vacua as before, similar to \Cref{eq:Gerochint}. Looking at the right-hand side of \Cref{eq:extendedboundaryterms}, it is natural to propose that the divergence-free set of tensors will be the conjugate partners of modes containing $\text{sub}^{n>1}$-leading soft news so that a functional derivative with respect to the corresponding conjugate mode will give a $\text{sub}^n$-leading soft insertion. This insertion is equivalent to the linear term in the renormalized higher spin charges proposed by Freidel, Pranzetti, and Raclariu\cite{Freidel:2021ytz}\footnote{It will require further analysis to have a first principle derivation of the $\text{sub}^n$-leading soft theorems from the on-shell action including the non-trivial right-hand side of the expression. One can not rely on the soft insertion to reproduce the corresponding soft factors since it is now well known that the RHS contains a non-universal piece for $n \geq 2$ and a term that spoils the factorization for $n\geq 3$\cite{Hamada:2018vrw,Laddha:2017ygw}. See Equation 4.2 in \cite{Akhtar:2024lkk}.}. We expect these modes to be a higher symmetry version similar to existing superrotation modes. Note that we will need more general falloffs on the news for this infinite tower of subleading news modes to exist, i.e., 
\be\label{eq;falloffs2}
N_{AB} \sim O(u^{-n-\epsilon}), \qquad |u| \to \infty
\ee
for a $\text{sub}^{n-1}$-leading soft insertion. A few comments are in order. For the whole tower of soft insertions, the news will have to decay faster than any polynomial. In generic classical scattering, even if we begin with compact and smooth initial data on $\ms{I}^-$, the scattered data will typically spread across all $\ms{I}^+$. Damour\cite{Damour:1985cm} pointed out that, massive objects in gravitational theory do not decouple even at late times and keep radiating leading to an algebraic decay of $O(u^{-2})$ of the news tensor\footnote{The $O(u^{-2})$
 behavior corresponds to the so-called gravitational tail of the memory effect and manifests as an $O(\ln\omega)$ subleading soft emission\cite{Campiglia:2021bap}.}. The consensus is that the fall-offs, \Cref{eq;falloffs2} are overly restrictive in realizing an infinite tower of symmetries of classical gravitational scattering. So, an important question arises regarding whether these Schwarzian fall-offs on the news can be relaxed when we work at classical level (see \cite{Nagy:2024jua} for a discussion in the case of Yang-Mills). Moreover, the tree-level subleading soft theorems gets loop corrections and lead to logarithmic terms in the soft expansion of S matrix amplitudes. Hence, it is expected that the loop \(w_{1+\infty}\) tree level symmetries proposed by Strominger \cite{Strominger:2021mtt} will get modified by loop effects. Secondly, no restriction prevents us from considering any form of
$u$ dependence, rather than restricting ourselves to polynomials, to define the set of tensors in \Cref{eq:generalizedGeroch}. The choice of polynomials was motivated by the fact that, in Fourier space, the $\text{sub}^n$-leading soft news has $\de_{\omega}^n$, which will make sense only when $n \in \mathbb Z^+$. Moreover, it is natural to choose the polynomial by observing the progression in which the constant shear mode, the coefficient of $u^0$ term, and $T_{AB}$, the coefficient of the linear in $u$ term, serve as the conjugate modes to memory and subleading memory, respectively.
\section{Discussion}We have demonstrated in this paper that the corner ambiguities in the boundary action at the null infinity of asymptotically flat spacetimes can be fixed by demanding that the (exponential of) on-shell boundary action factorizes into the Weinberg soft factor and the hard amplitude. We followed the analysis of Lehner, Myers, Poisson, and Sorkin to construct the boundary action at null infinity when the non-trivial memory mode is turned on. The corner ambiguities arise from the residual conformal freedom, $g'_{ab}=\gamma^2 g_{ab}$, and $\gamma \hateq 1 $. Working in a non-Bondi frame, we get an extra term in the boundary action that comes from the covariantization of the flux. This extra term can be written in terms of the subleading soft news smeared with the (STF) Geroch tensor and the vacuum mass aspect. We then extended the BMS group to allow (holomorphic) superrotations. A functional derivative of the $\mc{S}$ matrix with respect to the (STF) Geroch tensor gave us a subleading soft graviton insertion allowing us to derive the tree-level subleading soft graviton theorem from the AFS approach. The extension of our work to the full eBMS and gBMS requires further investigation. In both of these cases, complications arise due to the necessity of treating the boundary metric $q_{AB}$ as a dynamical variable on the phase space. Some unwanted and nonlocal terms were noticed by Kraus and Myers\cite{Kraus:2024gso} in an attempt to derive the subleading soft photon theorem for the massless QED case. A possible relationship with their work remains to be understood. By considering a generalization of the Geroch tensor to include a polynomial in $u$ with positive integer powers and coefficients given by a set of divergence-free symmetric traceless tensors on the celestial sphere (away from the punctures), the on-shell action leads to an infinite tower of $\text{sub}^n$-leading soft news terms. It is natural to propose that the set of tensors are the Goldstone modes forming a conjugate pair with the modes containing $\text{sub}^n$-leading soft news on (yet to be discovered) phase space. The presence of these terms hints towards a realization of the recently proposed infinite tower of soft graviton symmetries within the AFS framework\cite{Guevara:2021abz,Strominger:2021mtt,Freidel:2021ytz}.

 In \cite{Kim:2023qbl}, it was demonstrated that the AFS approach to the $\mc{S}$ matrix is equivalent to the LSZ prescription. If the AFS formulation of the tree-level $\mc{S}$ matrix stands correct, all classical symmetries should naturally emerge. We were able to demonstrate that the leading and subleading symmetries are manifest in the AFS approach based on the existing framework of the (extended) BMS group. The relationship between the symmetry interpretation of the infinite tower of soft insertions in the AFS approach and the higher spin symmetries governed by the $w_{1+\infty}$ algebra remains an open question and requires further investigation. This would require the construction of a phase space and the identification of charges corresponding to the $\text{sub}^n$-leading soft theorems, with an exploration of their connection to the existing works\cite{Freidel:2021ytz,Geiller:2024bgf}. The subleading soft factor in gravity and gauge theory receives quantum corrections due to long-range infrared effects, leading to new soft theorems with logarithmic dependence\cite{He:2014bga,Sahoo:2018lxl,Agrawal:2023zea}. A natural step forward would be to go beyond tree-level in the AFS path integral approach to the $\mc{S}$ matrix and derive these log soft theorems. We hope to explore these questions in future work.
\section*{Acknowledgements}
SU is very grateful to Alok Laddha for suggesting this problem, for numerous discussions, and for help with the draft. SU especially thanks Nishant Gupta and Adarsh Sudhakar for initial collaboration and discussions. SU wishes to thank Kartik Prabhu for several fruitful discussions.
\appendix
\section{Preliminaries}\label{sec:prelim}
 The metric of asymptotically flat physical spacetimes at $\ms{I^+}$ in the Bondi gauge with coordinates \((u,r,x^A)\) can be written as
 \be\label{Bondi gauge}
 d\hat{s}^2=-\frac{V}{r}e^{2\beta} du^2-2e^{2\beta} du\,dr+r^2 \hat h_{AB}(dx^A-U^A\,du)(dx^B-U^B\,du).
\ee
where, we have the asymptotic expansions
\be\label{eq:phys-BS-exp}
    \frac{V}{r}& = 1-\frac{2m_B}{r} + O(1/r^2)\,, \\
    \beta & = - \frac{1}{32 r^{2}} C_{AB}C^{AB} + O(1/r^3)\,, \\
    U^A & = -\frac{1}{2 r^{2}} D_{B} C^{AB} + O(1/r^3)\,, \\
   \hat h_{AB} & = q_{AB} +\frac{1}{r} C_{AB} +\frac{1}{4r^2} q_{AB} C_{CD} C^{CD} + O(1/r^3)\,.
\ee
Here $q_{AB}$ is the $u$-independent boundary metric(which is taken to be the unit sphere metric $\gamma_{AB}$ in most of the paper, see \Cref{sec:celestial} for a discussion on the conformal celestial metric) and $D^A$ is the covariant derivative with respect to this metric. We will work in the stereographical coordinates $x^A=(z,\bar{z})$. The fields $m_B$, $C_{AB}$ are functions of $(u,z,\bar{z})$ and regarded as tensors on $\ms{I^+}$. $m_B$ is called the Bondi mass aspect and the field $C_{AB}$ is the $O(r)$ term in the metric expansion of $\hat g_{AB}$. $C_{AB}$ is a symmetric, traceless tensor (STF) with two independent degrees of freedom. Further, it can be split into a time-dependent and a constant part\footnote{$C_{AB}$ is related to the asymptotic shear as $\text{STF} \, \nabla_a l_b \hateq -\frac{1}{2}C_{AB}$ \cite{Grant:2021sxk}.}
\be\label{eq:shear}
C_{AB}(u,z,\bar{z})=O(u^{-\epsilon})+\hat{C}_{AB}(z,\bar{z}), \qquad u \to \pm \infty
\ee
We further define
\be
N_{AB}=-\partial_{u}C_{AB}, \qquad \overset{0}{N}_{AB}=-\int_{-\infty}^{\infty}du N_{AB}=-2D_A D_B \overset{0}{N}
\ee
$N_{AB}$ is the Bondi News tensor and characterizes the gravitational radiation in the asymptotic region of the spacetime. $\overset{0}{N}_{AB}$ is the gravitational memory or the leading soft news. It can be written as 
\be
\overset{0}{N}=C_+-C_-
\ee
where $C_{\pm}$ is the value of the supertranslation field at $u \to \pm \infty$. The constant shear mode can also be written in a similar way
\be
\hat{C}_{AB}=-2D_A D_B \mc{C}, \qquad \mc{C}=\frac{1}{2}(C_+ + C_-)
\ee
This is sometimes called the Christodoulou-Klainermann (CK) condition. Note that this choice is not unique. The $\text{sub}^n$-leading news is defined as
\be
\overset{n}{N}_{AB}=-\int_{-\infty}^{\infty} du\, u^n N_{AB}, \qquad n \in \mathbb Z^+
\ee
The Fourier transform of the leading memory news goes as $\frac{1}{\omega}$ near $\omega=0$. This is what shows up as a pole in the Weinberg soft factor\cite{Ashtekar:2018lor,Prabhu:2022zcr}. The subleading soft graviton carries a prefactor $(1+\omega \de_{\omega})$ which projects out this pole thus eliminating the leading divergence and leaves the finite universal subleading piece\cite{Strominger:2017zoo}.
The inverse non-zero metric components are
\be
\hat{g}^{rr}=\frac{V}{r}e^{-2\beta}, \qquad \hat{g}^{ur}=-e^{-2\beta}, \qquad \hat{g}^{rA}=-U^Ae^{-2\beta}, \qquad \hat{g}^{AB}=\frac{1}{r^2}\hat h^{AB}
\ee
The asymptotic symmetry group at $\ms{I^+}$ as proposed in the original work\cite{Bondi:1962px} is the BMS group consisting of supertranslations and the Lorentz transformations. The vector field generating supertranslations at $\ms{I^+}$ is parametrized by a function $f(z,\bar{z})$ on the sphere
\be
\xi(f)&=f\partial_{u}+\left[
\frac{-D^{A}f}{r}+\frac{1}{2}\frac{C^{AB}D_{B}f}{r^2}\right]\partial_{A}\\
& \qquad\quad +\left[\frac{1}{2}D_{A}D^{A}f+\frac{-\frac{1}{2}D_{A}fD_{B}C^{AB}-\frac{1}{4}C^{AB}D_{A}D_{B}f}{r}\right]\partial_{r}+...
\ee
or in the conformal Bondi coordinates that we work in
\be
\xi(f) &= f n^a + (-\Omega\, D^A f + \Omega^2 \,C^{AB} D_B f)\partial_A - \frac{1}{2}\Omega^2 \,D_A D^A f \partial_\Omega + O(\Omega^3)
\ee
The action of supertranslations on the Bondi mass aspect and shear is 
\be\label{eq:mbfuture}
\Lie_{f}m_{B}=f\de_{u}m_{B}-\frac{1}{4}\left[N^{AB}D_{A}D_{B}f+2D_{A}N^{AB}D_{B}f\right] 
\ee
\be\label{eq:futureshear}
\Lie_f C_{AB}=f\de_u C_{AB}-2D_AD_Bf
\ee
where we use \(D_A D_B f\) as the shorthand notation for the trace-free part of this quantity throughout this paper. Note that the variation of the constant shear is $-2D_AD_Bf$. The $uu$ Einstein equation gives the constraint on the mass aspect
\be\label{eq:massaspectpositive}
\de_{u}m_{B}=-\frac{1}{4}D^{A}D^{B}N_{AB}-T_{uu}
\ee
where 
\be
T_{uu}=\frac{1}{8}N_{AB}N^{AB}+4\pi G\lim_{r \rightarrow \infty}\left(r^2
T_{uu}^{M}\right)
\label{eq:stress tensor1}
\ee
is the stress-energy tensor. For the energy flux to be finite across null infinity, the news tensor must decay to 0 at the boundary of $\ms{I^+}$
\be
N_{AB}= O(u^{-1-\epsilon}), \qquad |u| \rightarrow \infty
\ee
where $\epsilon >0$. Note that we will work with $\epsilon \neq 0$ in this paper to avoid log terms in the metric expansion. Using these fall-offs, from \Cref{eq:futureshear}, it can be easily seen that
\be
\Lie_f N_{AB}=f\de_u N_{AB}, \qquad \Lie_f \overset{0}{N}_{AB}=0
\ee
Along the same lines, we can look at the behavior of the metric at $\ms{I^-}$. The Bondi metric in advanced coordinates $(v,r,x^A)$ has the metric expansion at $\ms{I^-}$
\be
	d\hat{s}^2 &= -dv^2+ 2 \left(1-\frac{1}{16r^2}C_{AB}C^{AB}\right)\,dv\,dr + 2r^2\gamma_{AB}\,dx^A\,dx^B   \\
	& \qquad + \frac{2m_B}{r}dv^2 + \left( r C_{AB} \,dx^A\,dx^B  -   D^B C_{BA} \,dv\,dx^A \right)+.. \,,
\ee
 The supertranslation vector field at $\ms{I^-}$ is
\be
\eta\left(f\right)&=f\de_{v}+\left[
\frac{D^{A}f}{r}-\frac{1}{2}\frac{C^{AB}D_{B}f}{r^2}\right]\partial_{A}\\
& \qquad\quad +\left[-\frac{1}{2}D_{A}D^{A}f+\frac{\frac{1}{2}D_{A}fD_{B}C^{AB}+\frac{1}{4}C^{AB}D_{A}D_{B}f}{r}\right]\partial_{r}+...
\ee
Action of supertranslations on the Bondi mass aspect and shear is
\be\label{eq:mbpast}
\Lie_{f}m_{B}=f\de_{v}m_{B}+\frac{1}{4}\left[N^{AB}D_{A}D_{B}f+2D_{A}N^{AB}D_{B}f\right] 
\ee
\be\label{eq:pastshear}
\Lie_f C_{AB}=f\de_v C_{AB}+2D_AD_Bf
\ee
And from the $vv$ field equations, the constraint equation for the mass aspect is
\be\label{eq:massaspectnegative}
\de_{v}m_{B}= \frac{1}{4}D^{A}D^{B}N_{AB}+T_{vv}
\ee
The problem of classical gravitational scattering is concerned with finding a map from the initial Cauchy data on $\ms{I^-}$ to the final data on $\ms{I^+}$. To find such a map, it is important to observe what goes on at spatial infinity $i^0$ and we need to understand how to match the initial data in $\ms{I^-_+}$ to the final data at $\ms{I^+_-}$. For that we focus on Christodoulou-Klainermann(CK) spacetimes. CK showed nice enough initial data that decays sufficiently rapidly at $i^0$ so that the mapping corresponds to a smooth geodesically complete solution. For CK spaces the news tensor have more relaxed fall offs
\be
N_{AB}= O(u^{-3/2}), \qquad |u|\rightarrow \infty
\ee
or faster. We will not need these fall-offs for our analysis. $O(u^{-1-\epsilon})$ fall-offs of the news ensure that the flux of the radiation across $\ms I$ is finite and symplectic form \Cref{eq:Symplecticform} is well defined on the phase space. The vanishing of $N_{AB}$ at the endpoints of null infinity implies\footnote{Usually $\delta_f=\Lie_f$ for  covariant quantities. See \cite{Freidel:2021fxf, Odak:2022ndm} and the references therein for a discussion on anomalies for field dependent diffeomorphisms.}
\be
\Lie_f C_{AB}|_{\ms{I^{+}_{\pm}}}=-2D_AD_Bf
\ee
and
\be
\Lie_f C_{AB}|_{\ms{I^{-}_{\pm}}}=2D_AD_Bf
\ee
The matching condition we get across $i^0$ is
\be
C_{AB}|_{\ms{I^{+}_{-}}}= -C_{AB}|_{\ms{I^{-}_{+}}}
\ee
where the antipodal matching of the coordinates is implicit. For the mass aspect, it is
\be
m_B(u=-\infty)=m_B(v=+\infty)
\ee
The matching conditions for a large class of spacetimes were proved in \cite{Prabhu:2019fsp,Capone:2022gme}.
\subsection{Covariant phase space and CFS modification}\label{sec:CFS}
Let $L$ be a diffeomorphism-invariant Lagrangian $d$-form. Its variation is
\begin{equation}
\delta L = E \cdot \delta \phi + d\theta(\phi;\delta\phi),
\end{equation}
where $\theta$ is the presymplectic potential $(d-1)$-form. The standard (Wald) symplectic current pullback is
\begin{equation}
\pb \omega_{\text{Wald}}(\phi;\delta_1,\delta_2)
=
\delta_1 \Theta(\phi;\delta_2) - \delta_2 \Theta(\phi;\delta_1).
\end{equation}
where \(\Theta\) is the boundary symplectic potential. Chandrasekaran--Speranza (CS) \cite{Chandrasekaran:2020wwn} define the pullback of $\theta$ to a hypersurface $\Sigma$, which modifies the boundary potential by a corner term:
\be
\pb \theta &= -\delta S_N + d\beta + \Theta \\
\Theta &\;\longrightarrow\; \Theta + \delta \beta,
\ee
where \(S_N\) is the boundary term/boundary Lagrangian, $\beta$ is a $(d-2)$-form corner term localized on future and past boundaries of null surface $\Sigma$. The resulting symplectic current is
\begin{equation}
\omega_{\text{CS}}
=
\omega_{\text{Wald}} + d\Big( \delta_2 \beta - \delta_1 \beta \Big).
\end{equation}
Integrating over a hypersurface $\Sigma$:
\begin{align}
\Omega_{\text{CS}}
&=
\int_{\Sigma} \omega_{\text{CS}} \\
&=
\int_{\Sigma} \omega_{\text{Wald}}
+
\int_{\partial \Sigma}
\Big( \delta_2 \beta - \delta_1 \beta \Big).
\end{align}
Thus,
\begin{equation}
\Omega_{\text{CS}}
=
\Omega_{\text{Wald}} + \Omega_{\text{corner}}.
\end{equation}
Let $\Sigma = \mathscr I^+$, with boundaries at $u=\pm\infty$ spheres. The Wald symplectic form reduces to the Ashtekar--Streubel form:
\begin{equation}
\Omega_{\text{AS}}
=
-\frac{1}{32\pi G}
\int_{\mathscr I^+} du\, d^2z \gamma_{z\bar z} \;
\delta N^{AB} \wedge \delta C_{AB}.
\end{equation}
The CS correction gives
\begin{equation}
\Omega_{\text{corner}}
=
\frac{1}{32\pi G}
\int_{ \mathscr I^+_{\pm}} d^2z\, \gamma_{z \bar z}
\left(\delta_2 \beta - \delta_1 \beta\right).
\end{equation}
Using the corner terms derived in the paper, 
the modified symplectic form is
\begin{equation}
\Omega_{\text{CS}}
=
-\frac{1}{32\pi G}
\left(\int_{\mathscr I^+} du\, d^2z \,\gamma_{z\bar z} \;
\delta N^{AB} \wedge \delta C_{AB}
+
\int_{S^2} d^2z \, \gamma_{z\bar z} \,
\delta(\overset{0}{N}_{AB}) \wedge \delta(\hat C^{AB})\right)
\end{equation}
This is the symplectic form which is consistent with the leading Weinberg soft theorem. In principle, one can try to find corner terms which will provide a further extension of the corner symplectic form to include subleading symmetries, but we do not do that here.
\subsection{Action of superrotations}
The vector fields generating superrotations at $\ms{I^+}$ is
\begin{align}
\begin{split}
\xi_{Y}&=\alpha\partial_{u}+\left(Y^{A}-\frac{1}{r}D^A\,\alpha+\frac{1}{2r^2}C^{AB}D_B\,\alpha+O(r^{-3})\right)\partial_{A}\\
&+\left(-\frac{r}{2}D_{A}Y^{A}+\frac{1}{2}D^2\alpha+\frac{1}{r}\left(-\frac{1}{2}C^{AB}D_AD_B\,\alpha-\frac{1}{4}C^{AB}D_AD_B\,\alpha\right)+O(r^{-2})\right)\partial_{r}
\end{split}
\end{align}
where $\alpha=\frac{u}{2}D_AY^A$ and $Y^{A}$ is a local CKV/smooth vector field on the sphere. The variation of various field quantities under local CKV/$\text{Diff}(S^2)$ can be found in \cite{Compere:2018ylh}. 
We note down the variations of relevant quantities for completeness,
\be
\delta_Y\hat{C}_{AB}=\left(\Lie_Y-\frac{1}{2}D_C Y^C\right) \hat{C}_{AB}
\ee
OR
\be
\delta_Y \mc{C}= \left(\Lie_Y-\frac{1}{2}D_CY^C\right)\mc{C}
\ee
For the Geroch tensor we have
\be
\delta_Y T_{AB}= \Lie_Y T_{AB}- D_A D_B D_C Y^C
\ee
The variation of the subleading soft news is
\be
\delta_Y \overset{1}{N}_{AB}= \left(\Lie_Y- D_C Y^C\right)\overset{1}{N}_{AB}
\ee
$\text{Diff}(S^2)$ acts on the metric $q_{AB}$ in such a way that $\delta_Y \sqrt{q}=0$. The subleading soft graviton theorem follows from the Ward identity of the superrotation charge which physically corresponds to the conservation of angular momentum at each angle \cite{Campiglia:2014yka,Kapec:2014opa}\footnote{The charges are agnostic about whether the vector field is a local CKV or a smooth diffeomorphism on the sphere. A matching condition has also been conjectured for the angular momentum $N_A$ but it diverges logarithmically because of log corrections to the subleading soft factor at one loop\cite{Compere:2023qoa}. }.
\section{Conformal Gaussian null coordinates/Newman-Unti gauge}\label{sec:NewmanUnti}
Instead of demanding that $\Omega$ be a ``radial'' coordinate, we can choose the conformal factor such that the normal $l^a\hateq-\frac{\de}{\de\Omega}$ becomes the generator of null geodesics transverse to $\ms{I}$ so that $l^b\nabla_b l^a=0$. Additionally, we can select a new frame($\Omega'=\omega\,\Omega$) in which the expansion of the auxiliary normal vanishes and $\Omega'$ serves as an affine parameter along the transverse null geodesics.  For simplicity, we will omit the``prim'' notation for $\Omega'$ going forward.

The most general form of the unphysical metric in these coordinates is given by\cite{Grant:2021sxk}
\be\label{eq:unphys-CGN-metric}
    ds^2 & = 2 du (d\Omega - \alpha du - \beta_A dx^A) + h_{AB} \,dx^A \,dx^B\,,
\ee
Imposing the Bondi condition \(\nabla_a n_b \hateq 0\) leads to the following conditions
\be
    \alpha = O(\Omega^2) \eqsp \beta_A = O(\Omega^2)\,.
\ee
To write an expression for the physical metric define \(\lambda \defn \Omega^{-1}\) so that in the coordinates \((\lambda, u, x^A)\) the physical metric \(\hat g_{ab} = \Omega^{-2} g_{ab} = \lambda^2 g_{ab}\) has the components
\be\label{eq:phys-metric-GNC}
    \hat{g}_{\lambda\lambda} & =0 \eqsp
    \hat{g}_{A\lambda}=0 \eqsp 
    \hat{g}_{u\lambda}=-1 \eqsp \\
    \hat{g}_{uu} & =-1 - \frac{1}{\lambda} \mc{P}+ O(1/\lambda^2) \eqsp \\
    \hat{g}_{uA} & =\frac{1}{2} {D}^{B}C_{AB} + O(1/\lambda) \eqsp \\
    \hat{g}_{AB} & = \lambda^{2} q_{AB} + \lambda C_{AB}+ \frac{1}{8} q_{AB} C_{CD} C^{CD} + O(1/\lambda) \,. 
\ee
where $\mc P$ is the Weyl scalars \(\Omega^{-1} C_{nlnl}\)
\be
\mc{P}  = 2 \alpha^{(3)}
\ee
The vector field
\be
    \hat l^a \equiv \frac{\de}{\de{\lambda}} = - \Omega^2 \frac{\de}{\de{\Omega}}\,,
\ee
generates outgoing null geodesics that are affinely parametrized with respect to the physical metric \(\hat g_{ab}\) with \(\lambda\) serving as the affine parameter. This gives the physical metric in the coordinates used in the \emph{affine-null} form used in the Newman-Unti coordinates \cite{Newman:1962cia}. We first compute
\begin{align}
n_a n^a = g^{\Omega\Omega}
= 2\alpha + h^{AB}\beta_A \beta_B
= \Omega^2 + 2\Omega^3 \alpha^{(3)} + O(\Omega^4)\,.
\end{align}
The inaffinity is given by
\begin{align}
\kappa = -\tfrac{1}{2} \Lie_l (n_a n^a)\,, \qquad l^a = -\partial_\Omega\,,
\end{align}
which yields
\begin{align}
\kappa = \Omega + 3\Omega^2 \alpha^{(3)} + O(\Omega^3)\,.
\end{align}
The expansion is
\begin{align}
\theta = \tfrac{1}{2} h^{AB} \Lie_n h_{AB}\,.
\end{align}
for which we obtain
\begin{align}
\theta 
= -\tfrac{1}{8}\Omega^2 \partial_u \bigl(C_{AB} C^{AB}\bigr) 
+ O(\Omega^3)\,.
\end{align}
The physical quantities are related by
\begin{align}
\hat\theta &= \Omega^2 \theta - 2\Omega\, n_a n^a\,, \\
\hat\kappa &= \Omega^2 \kappa + \Omega\, n_a n^a\,.
\end{align}
Substituting the above expansions, we find
\begin{align}
\hat\theta &= -2\Omega^3 
- 4\Omega^4 \alpha^{(3)}
- \tfrac{1}{8}\Omega^4 \partial_u (C_{AB}C^{AB})
+ O(\Omega^5)\,, \\
\hat\kappa &= 2\Omega^3 
+ 5\Omega^4 \alpha^{(3)}
+ O(\Omega^5)\,.
\end{align}
Therefore, the combination of interest is
\begin{align}
-2(\hat\theta + \hat\kappa)
= -2\Omega^4 \left(
\alpha^{(3)} - \tfrac{1}{8}\partial_u (C_{AB}C^{AB})
\right)
+ O(\Omega^5)\,.
\end{align}
Using the identification \(\mathcal P = 2\alpha^{(3)}\), this can be written as
\begin{align}
-2(\hat\theta + \hat\kappa)
= -\Omega^4 \left(
\mathcal P - \tfrac{1}{4}\partial_u (C_{AB}C^{AB})
\right)
+ O(\Omega^5)\,.
\end{align}
The boundary term on \(\ms I^+\) is
\be
S_k \hateq -\frac{1}{16\pi G} \int_{\ms I^+} du \, d^2\, \gamma_{z\bar{z}} \left(
\mathcal P - \tfrac{1}{4}\partial_u (C_{AB}C^{AB})
\right)
\ee
One can use corner ambiguities to fix a supertranslation invariant action in these coordinates.

\bibliographystyle{JHEP}
\bibliography{sample}  
\appendix

\end{document}